\documentstyle[twocolumn,aps,epsfig,floats]{revtex}
\begin{document}
\draft

\title{The L\'evy diffusion as an effect of sporadic
randomness}
\author{Mauro Bologna $^{1}$, Paolo Grigolini$^{1,2,3}$ and Juri
Riccardi$^{4}$}
\address{$^{1}$Center for Nonlinear Science, University of North Texas,\\
P.O. Box 5368, Denton, Texas 76203 }
\address{$^{2}$Istituto di Biofisica del Consiglio Nazionale delle\\
Ricerche, Via San Lorenzo 26, 56127 Pisa, Italy }
\address{$^{3}$Dipartimento di Fisica dell'Universit\`{a} di Pisa,
Piazza Torricelli 2, 56127 Pisa, Italy }
\address{$^{4}$ENEL SpA Struttura Ricerca-Area Generazione Via A. Pisano 120, 56100,
Pisa, Italy}
\date{\today}
\maketitle

\begin{abstract}
The
 L\'{e}vy diffusion processes are a form of non ordinary
 statistical mechanics resting, however, on the conventional
 Markov property. As a consequence of this, their dynamic derivation is
 possible provided that (i)
 a source of randomness is present in the corresponding microscopic
 dynamics and (ii) that the consequent process of memory erasure is
 properly taken into account by the theoretical treatment.
\end{abstract}


\section{Introduction}

The theoretical treatment of anomalous diffusion, namely diffusion
processes either faster or slower than ordinary Brownian diffusion,
is an active field of
research.
A well known case of superdiffusion is given by
the diffusion processes
 of L\'evy type\cite{LEVY}.
 The reader can find an exhaustive discussion of the recent literature
 on this subject in excellent review papers\cite{MONTROLL,bouchaud,SZK,KZS}. However,
we would like to draw the reader attention also on some key papers
of the literature on this subject
\cite{KBS,SWK,BZK,seshadriwest,westseshadri,peseckis,fogedby,honkonen,cp,metzler,metzler1,KZ,ZK}.
We plan to adopt a dynamic \cite{MANNELLA,ALLEGRO} rather than a
probabilistic
approach\cite{KBS,SWK,BZK,seshadriwest,westseshadri,peseckis,fogedby,honkonen,cp,metzler,metzler1,KZ,ZK}.
To make the significance of this purpose more transparent it is
convenient to compare what we mean by \emph{dynamic} to the
conventional probabilistic treatment, either resting on the
L\'{e}vy flight or the L\'{e}vy walk method.

\subsection{L\'{e}vy flight and L\'{e}vy walk}
Both the L\'{e}vy flight and the L\'{e}vy walk method are based on
a totally probabilistic treatment. The L\'{e}vy flight method is
based on the assumption that at regular time intervals a space
transition of arbitrarily large intensity might take place. With
the L\'{e}vy walk, on the contrary, the jumps over larger
distances take place in larger times. This property make these
processes non-Markovian and consequently the derivation of
L\'{e}vy diffusion more delicate than from within the L\'{e}vy
flight perspective.This is easily realized, for instance, by using
the continuous time random walk formalism\cite{KBS,cp} and
expressing the time evolution for the probability that the
particle is at a given space location at a given time by means of
the equivalent generalized master equation, see
also\cite{metzler,metzler1}. It is then easily seen that the case
where the waiting time distribution is characterized by a finite
time scale yields immediately a Markov process in the long-time
limit, and the anomalous diffusion properties only depend on the
long-range nature of the displacement per step distribution. In
Section II of this paper we shall study a physical condition of
the same kind. Some special attention has to be devoted therefore
to the L\'{e}vy walk condition, since it shares with the dynamic
approach (see Subsection 1C for a more precise definition of this
approach) a long-time memory, which has to be properly erased to
establish in the long-time limit the conditions for L\'{e}vy
statistics.

 \subsection{From the probabilistic to the dynamic approach}
 The dynamic approach, whose precise meaning will be discussed
 in Section 1C,  is still somewhat obscure due to
 some conflicting aspects of the recent theoretical
 derivations used to realize  this
 goal. This is so, in spite of the fact that the
 explicit adoption of
 techniques derived by the random
 walk literature yields
 a satisfactory  derivation of the processes
 of L\'{e}vy diffusion from within the
 theoretical framework of
 probabilistic treatments\cite{SWK},\cite{BZK},\cite{SZK},\cite{KZS}.
  The purpose of this paper is that of affording a unified
 perspective with no internal contradictions.
 Here we limit ourselves to point out some aspects of the L\'{e}vy
 walk method which must be retained by the dynamic approach.

First of all, as made clear
by the work of Klafter and Zumofen \cite{KZ} and
Zumofen ad Klafter \cite{ZK} (see also the report
of Klafter, Zumofen and Schlesinger \cite{KZS}),
 we have to point out that
the process of L\'evy diffusion can be derived from within a
dynamic perspective if the so called L\'evy walk view is adopted.
This means a trajectory  moving with constant velocity along a
straight line for an extended time and from time to time making
abrupt direction changes.  The time of sojourn  in one of these
straight paths is characterized by the probability density
function
\begin{equation} \label{eq1}
\psi(t)=\frac{(\mu-1)\,T^{\mu-1}}{(T+t)^\mu}\,\,,
\end{equation}
where $T/(\mu-2)$ denotes the mean waiting time.
The renormalization group method, as
illustrated by Zaslavsky \cite{ZAS}, affords a reliable way of  fixing
the time asymptotic form of (\ref{eq1}), and, notably, the
power index $\mu$ in terms of the rescaling properties of the
fractal region at the border between chaotic sea and stability islands.
The theorem of Kac \cite{KAC} ensures that the first moment of
$\psi(t)$ is finite. This important theorem refers to
the distribution of the Poincar\'{e} recurrence times under the crucial
condition that the system under study is ergodic.
 Zaslavsky\cite{ZAS} noticed
that when a stability island is imbedded within the chaotic
sea, the distribution of Poincar\'{e} recurrence times becomes
equivalent to the distribution of the times of sojourn
at the border between chaotic sea and stability island. This is so
because a trajectory moving from a given small portion of the
chaotic sea through a fast diffusion process arrives
at the fractal region and it sticks to it for
an extended time before returning to the departure region.Thus, in
a full accordance with the Kac theorem, the renormalization
group theory  yields $\mu>2$,
thereby insuring that the first moment of this
distributions is finite, and consequently that also the
distribution of the Poincar\'{e} recurrence times is finite.

We note that all this supports the asymptotic form
of (\ref{eq1}) leaving, though, the impression
that its short-time structure is
arbitrary. It is not so. As proved in a recent work, \cite{ANNA},
the whole structure of (\ref{eq1}) is dictated  by the principle of entropy
maximization provided that the entropy used is
that of Tsallis \cite{COSTA} rather than the conventional
Gibbs-Shannon entropy. Note that the explicit form of (\ref{eq1}) that
we are using, is fixed, of course, by both
the normalization condition and the condition
that the first moment is finite and that its value
is $T/\,(\mu-2)$. In conclusion, the form of Eq.(\ref{eq1}) is
a unique analytical expression determined by the joint use of
dynamics, renormalization group technique, and entropy.

We hope that with no sacrifice of the most important ingredients
behind the dynamic derivation of L\'evy diffusion process, we can
restrict our investigation to the one-dimensional case. In this
condition the role of deterministic and dynamical generator of the
L\'evy diffusion can be properly played by the intermittent map
\cite{IM} used by Zumofen and Klafter \cite{ZK} and by Klafter and
Zumofen \cite{KZ}. This is a map with the same algorithmic
complexity  as the Manneville map \cite{MP}, the complexity of
which  has been studied by Gaspard and Wang \cite{GW} by means of
the Kolmogorov-Sinai entropy.

In the Hamiltonian model of Zavslasky the
derivation of the diffusion processes of L\'evy kind
rests on the microcanonical conditions. This
means that the kinetic energy of the flight process
is fixed. Consequently, the one-dimensional version of this
Hamiltonian perspective yields
the important property that only two velocity states exist,
one with velocity $W$ and one with velocity $-W$.
As a consequence of the one-dimensional
assumption, therefore, we are allowed to use the key
relation\cite{IM}
\begin{equation}\label{eq2}
\Phi_\xi(t)=\frac{\mu-2}{T}\int_{t}^{\infty}(t'-t)\psi(t')\,dt'\,\,,
\end{equation}
which is equivalent to
\begin{equation}
\psi(t) = \frac{T}{\mu-2} \frac{d^{2}}{dt^{2}}\Phi_{\xi}(t).
\label{psiaseconderivative}
\end{equation}
Equations (\ref{eq2}) and (\ref{psiaseconderivative}) relate to
one another the physical properties $\psi(t)$ and $\Phi_{\xi}(t)$.
The former property , $\psi(t)$, is the probability density
function  of sojourn times, which, as earlier stressed, has an
inverse power law form (see Eq.(\ref{eq1})); the latter,
$\Phi_\xi(t)$, is the stationary correlation function of the
dichotomous variable $\xi$, playing the role of a velocity with
only two possible values, $W$ and $-W$. The function
$\Phi_{\xi}(t)$ is determined by the statistical properties of the
velocity of the paths moving with constant velocity and without
changing direction. We note that Eq. (\ref{psiaseconderivative})
establishes that $\psi(t)$ is proportional to the second-order
time derivative of the function $\Phi_{\xi}(t)$, thereby implying,
as a consequence of Eq. (\ref{eq1}), that for $t\rightarrow
\infty$ the decay of $\Phi_{\xi}(t)$ is proportional to
$1/t^{\beta}$ with $\beta = \mu -2$. The region of interest for us
is that where the first moment of $\psi(t)$ is finite (so as to
fit the Kac theorem) and the second moment is divergent so as to
prevent the system from falling in the attraction basin of the
central limit theorem. Consequently, we restrict our analysis to
the interval $0< \beta <1$. Note that (\ref{eq2}) is exact (see
\cite{IM}) if the assumption is made that the time interval
between the transition from one to the other velocity state is
instantaneous. To help the reader to understand the main
conclusion of this paper we have also to make another preliminary
remark. The correlation function $\Phi_{\xi}(t)$ is a stationary
property\cite{MANNELLA}, implying the existence of an invariant,
or equilibrium, distribution. A genuinely dynamic approach to  the
L\'{e}vy processes, consistent with the ergodic assumption behind
the Kac theorem, implies that this equilibrium distribution is
established by  a single trajectory, provided that this trajectory
runs for an unlimited amount of time. The lack of a finite
microscopic time scale makes this condition difficult to realize
in practice, and it is probably one of the sources of the
conflicting views that will be discussed in this paper.

\subsection{The dynamic approach}
The general program of the dynamic approach to statistical
mechanics is illustrated in a series of recent
papers\cite{HHPP87,PP96,J95,BMWG95}. We are very close to the
program of Ref.\cite{BMWG95}. The ambitious purpose of these
authors is to derive an important equation as the Fokker-Planck
equation without using any statistical assumption whatsoever, so
as to reverse the ordinary path from thermodynamics to statistical
mechanics. In other words, the path to follows moves from dynamics
and reaches the level of statistical mechanics using only
deterministic randomness with no recourse to thermodynamics, this
being the last step, stemming from the dynamically generated
statistical equilibrium distributions.

The authors of \cite{ALLEGRO} adopted the same perspective to move
from dynamics to L\'{e}vy statistics. The authors of Ref.\cite{ALLEGRO}
found that the density
distribution $\sigma(x,t)$ of the variable $x$ driven by a process
described by Eq.(\ref{eq2}) obeys the
 equation of motion \cite{ALLEGRO}:
\begin{equation} \label{eq3}
\frac{\partial}{\partial\,t}\,\sigma(x,t)=\langle\,\xi^2\,\rangle\,
\int_{0}^{t}\Phi_\xi(t')\,\frac{\partial^2}{\partial\,x^2}\,
\sigma(x, t-t')\,dt'   .
\end{equation}
Within the context of a dynamic approach
to the L\'{e}vy processes, this equation should
be given a special attention, since no explicit use of probabilistic
arguments was made to derive it\cite{ALLEGRO}.
However, no general solution of it is available, and
the emergence
of the L\'{e}vy diffusion out of it rests
on an approximation which has been questioned\cite{TC1,TC2}:
Different approximations to the solution of Eq.(\ref{eq3}) lead
to different statistical processes.
 The interested reader
is referred to the work of Ref.\cite{ALLEGRO} for the
the derivation of Eq.(\ref{eq3}). Here we limit ourselves to noticing
that this equation is exact under the condition that
the velocity variable is dichotomous and the initial distribution is a
Dirac delta centered at $x=0$.
Thus, there is an intimate relation between Eq.(\ref{eq2}) and
Eq.(\ref{eq3}).

\subsection{Purpose and outline of the paper}

The main purpose
of this paper is that
of stressing  that the intimate connection between Eq. (\ref{eq2}) and
Eq.(\ref{eq3})
does not leave room for a solution
implying a memory infinitely extended in time.
 In other words, we
want to prove that the adoption of a
Markovian perspective, although apparently incompatible
with the time convolution
of  Eq. (\ref{eq3}), is dictated by the steady action of the randomness
corresponding to the transition from the one to the other state of the
variable velocity.
Consequently the Markov structure emerging
from Eq.(\ref{eq3}) according to the prescriptions of
\cite{ALLEGRO} cannot be misled as an undue
approximation. This is rather an ingenuous  way of
establishing a physical condition fitting
the result of an earlier, and crucial, research work.
Gaspard and Wang \cite{GW} prove that in the long-time
limit the Manneville map becomes equivalent to
the Bernouilli shift map. This is a consequence
of the fact that the repeated exit and
re-entering into the laminar region results in a
memory erasure. As we shall see, this is the main reason why
the  final
state is as Markov as a genuine L\'evy process
must be.

In other words, we plan to make randomness emerge from the dynamic
approach, resting on
Eq.(\ref{eq3}), so as to render this dynamic approach
equivalent to the
L\'{e}vy walk perspective.
The purpose of this paper is to show that the Markov property
necessary to derive the process of L\'{e}vy diffusion is not
arbitrary, but rather corresponds intimately to the nature of the
dynamic process resulting in Eq.(\ref{eq3}). This is so because
Eq.(\ref{eq3}) implies the dichotomous nature of the variable
$\xi$. The condition where  for $t\rightarrow \infty$ the decay of
$\Phi_{\xi}(t)$
is proportional
to $1/t^{\beta}$ with $\beta = \mu -2$
means that the process is deterministic
for an overwhelming amount of time.
There exists an intimate equivalence between Eq.(\ref{eq2}), in this
physical condition, and intermittent maps. Randomness shows
up only when the trajectory reaches the border between laminar and
chaotic region\cite{MP}. At this crucial stage there are only two
possible events, either a jump from the original into the other laminar
region,
corresponding to a distinct velocity state, or the jump back to
the original laminar region, namely, the original velocity
state\cite{ZK}. At this stage dynamics are essentially
indistinguishable from the time evolution
of the Bernouilli shift map, whose connection with
thermodynamics and statistical mechanics has been recently clearly
illustrated by Zaslavsky\cite{ZAS}.
This means that randomness is a rare event and it is
in fact the reason why we have adopted the concept of
\emph{sporadic randomness}. The main purpose of this paper
is that of making a choice between two distinct ways of solving
Eq.(\ref{eq3}), based on the criterion that the right solution
must reflect this sporadic randomness.

The outline of this paper is as follows. In Section II we review the
arguments used in an earlier paper,\cite{TREFAN}, to derive a process
of L\'{e}vy diffusion by means of a generalized master equation. In
Section III, using the calculation illustrated in the Appendix, we
show that the same result is derived from a master equation which
looks like the Markov approximation of that of Section II. In Section
IV we review, in the light of the perspective established in this
paper, the method used in Ref.\cite{ALLEGRO} to assign to
Eq(.\ref{eq3}) a Markov structure. Some final conclusions are made
in Section V.

\section{The generalized master equation}
The first step of our approach rests  on the use of
the generalized master equation of Ref.\cite{KENKRE}. This
equation reads:
\begin{equation}  \label{generalized}
\frac{\partial}{\partial t}\sigma(x,t)=\int_{0}^{t}\,dt'
\int_{-\infty}^{\infty}\kappa(x-x',t-t')\sigma(x',t')\,dx' \,\,,
\end{equation}
where
\begin {equation}
\kappa(x,t) \equiv \pi(x,t) - \delta(x) \int_{-\infty}^{\infty}dx'
\pi(x',t)
\label{auxiliarytogeneralized}
\end{equation}
and $\pi(x,t)$ denotes the probability for the particle to make a
jump by a distance $x$ at time $t$.
This equation is very general and is expected to be compatible with
the description of highly non-Markovian processes like that
corresponding to Eq.(\ref{eq3}) with the non integrable correlation
function generated by Eqs.(\ref{eq1}) and (\ref{eq2}).
The intimate connection between these two equations will be discussed
in Section IV. Here we limit ourselves to stressing that the
asymptotic regime of $\sigma(x,t)$, as given by Eq.(\ref{generalized}),
can be studied without making explicitly the Markov approximation. In
fact, using the property that this equation is convoluted in both space
and time variables, we get for the Fourier-Laplace transform of $\sigma(x,t)$,
denoted by $\hat{\sigma}(k,s)$, the following expression:
\begin{equation}
\hat{\sigma}(k,s) = \frac{1}{s- \hat{\kappa}(k,s)},
\label{nonmarkovapproximation}
\end{equation}
where, of course, $\hat{\kappa} (k,s)$ denotes the Fourier-Laplace
transform of $\kappa(x,t)$. As pointed out in Section I, the
dynamic approach to L\'{e}vy statistics that we are considering is
related to the L\'{e}vy walk condition. This means that a
transition of length $|x|$ implies a time $t = |x|/W$. In
conclusion we are forced to make the following choice for
$\pi(x,t)$, with $t>0$:
\begin{equation}
\pi(x,t) = \psi(t) \delta(|x| - Wt).
\label{flight}
\end{equation}
The authors of Ref.\cite{TREFAN} studied the asymptotic  regime of
Eq.(\ref{nonmarkovapproximation}), supplemented by
Eq.(\ref{flight}),  searching  for the rescaling
condition
\begin{equation}
s \propto k^{\alpha},
\label{rescaling}
\end{equation}
with $\alpha > 1$. This is a reasonable assumption, since in
the asymptotic limit the second moment is known\cite{TREFAN} to yield
\begin{equation}\label{eq8}
x\approx t^{2\,H}
\end{equation}
with:
\begin{equation}\label{eq9}
H=1-\beta/2\,\,.
\end{equation}
On the other hand the rescaling of Eq.(\ref{rescaling}) suggests that
the condition
$\alpha = 1/H$ might apply, thereby resulting
in the property $\alpha >1$, which
is essential for the calculations aiming at establishing the exact
dependence of $\alpha$ on $\beta$.

With straightforward calculations it is shown\cite{TREFAN}
that
in the asymptotic limit Eq.(\ref{nonmarkovapproximation}) and
Eq.(\ref{flight}) yield:
\begin{equation}
\alpha = \beta + 1
\label{levyrescaling}
\end{equation}
and
\begin{equation}
\hat{\sigma}(k,s) = \frac{1}{s+ b|k|^{\alpha}},
\label{levystructure}
\end{equation}
with
\begin{equation}
b \equiv \Gamma(1-\alpha) \frac{(W\,T)^{\alpha}}{T}(\alpha -1)
cos(\frac{\alpha \pi}{2}). \label{defineb}
\end{equation}
Note that this is the Laplace transform of the following equation of
motion:
\begin{equation}
\frac{\partial
}{\partial\,t}\,\hat{\sigma}(k,t)=-b\vert\,k\,\vert^{\alpha}\hat
{\sigma}(k,t) .
\label{final}
\end{equation}
This means that the asymptotic regime of the generalized master
equation of Eq.(\ref{generalized}) is a process of diffusion with a
genuinely L\'{e}vy nature\cite{MONTROLL}.

We note that Eq.(\ref{levyrescaling}) means the
rescaling

\begin{equation}\label{eq10}
H=\frac{1}{1+\beta}\,\,,
\end{equation}
which is different from that of Eq.(\ref{eq9}). The difference
between the two rescalings is a fact of crucial importance
deserving proper comments. We note that the rescaling of
Eq.(\ref{eq9}) is somewhat ambiguous since it refers to the
dynamics of Eq.(\ref{nonmarkovapproximation}). As pointed out in
Ref.\cite{ALLEGRO}, as well as in the earlier work of Zumofen and
Klafter\cite{ZK} and Klafter and Zumofen\cite{KZ}, the diffusion
process described by Eq.(\ref{nonmarkovapproximation}) consists of
a central part and a propagation front signalled by two sharp
peaks.  At time $t$ a particle leaving the origin $x=0$ at $t=0$
cannot be found at a distance from the origin larger than $Wt$.
This has the effect of producing an accumulation of particles at
the front itself of the diffusion process, namely at $x = \pm W
t$. This is the origin of the two ballistic peaks of the
propagation front. At earlier times the initial distribution,
concentrated at $x=0$, splits into these two ballistic peaks and
the region between the two peaks is empty.  Due to the effect of
sporadic randomness, some trajectories leave the propagation front
and the population of the central part steadily increases in time,
while the peak intensity, proportional to the correlation function
$\Phi_{\xi}$ slowly decreases. Note that this means that the
diffusion process cannot be described by a single rescaling.  The
peaks of the propagation front rescale with $H = 1$, a fact
implying a diffusion faster than that predicted by the rescaling
of Eq.(\ref{eq9}). The rescaling of the central part is properly
expressed by Eq.(\ref{eq10}). The calculations leading to
Eq.(\ref{final}) refer to a physical condition where the intensity
of the ballistic peaks is negligible, so that the rescaling of
Eq.(\ref{eq10}) only reflects the diffusion properties of the
distribution central part. On the contrary, the rescaling of
Eq.(\ref{eq9}) is a sort of balance between the fast rescaling of
the propagation front and the rescaling of the central part of the
distribution $\sigma(t)$, which is in fact  slower than rescaling
of Eq.(\ref{eq10}). In conclusion, the discrepancy between the
rescaling of Eq.(\ref{eq9}) and the rescaling of Eq.(\ref{eq10})
reflects the fact the derivation of the L\'{e}vy process from
within the L\'{e}vy walk perspective rests on a deep conflict
between the dynamic properties still present within the L\'{e}vy
walk perspective  and a merely probabilistic treatment.

We want to make a further remark, concerning the derivation of a
L\'{e}vy process from the generalized master
equation of Eq.(\ref{generalized}).
We note that the L\'{e}vy processes are a form of
Markov statistics \cite{LEVY} thereby implying that the asymptotic
regime of the dynamic process described by Eq. (\ref{generalized})
involves a process of memory erasure. This paper is devoted to
proving that Eq.(\ref{eq3}), in spite of its
non-Markovian structure,  yields an asymptotic regime
whose statistics are determined by this process
of memory erasure.

\section{The Markov master equation}
As pointed out at the end of Section II, the processes of L\'{e}vy
diffusion are Markovian. Therefore, it is convenient to discuss
their derivation from a Markov master equation. This discussion
will serve the useful purpose of proving that the master equation
is, in a sense, a bridge between the dynamic treatment and the
diffusion regime. The latter is the subject of thermodynamical
arguments and the former rests, in the theoretical picture adopted
in this paper, on classical mechanics. Thus, the master equation
can also be regarded as an important bridge between mechanics and
thermodynamics. In the continuous representation the master
equation reads, see \cite{MONTROLL}, \cite{BEDEAUX} and
\cite{TREFAN}:
\begin{equation}\label{eq4}
\frac{\partial}{\partial\,t}\sigma(x,t)=\int_{-\infty}
^{\infty}K(x-x')\,\sigma(x',t)\,dx'\,\,,
\end{equation}
where:
\begin{equation}\label{eq5}
K(x)=\Pi(x)-\delta(x)\int_{-\infty}^{\infty}\Pi(x')dx'.
\end{equation}

As far as the transition probability $\Pi(x)$ is concerned, we
adopt the result of the entropic analysis of Ref. \cite{ANNA}.
Thus we write:
\begin{equation}\label{eq6}
\Pi(x)=\frac{1}{T}\frac{1}{W}\,\psi\Big(\frac{\vert\,x\,\vert}{W}\Big)=\frac{(\mu-1)\,T
^{\mu-1} \,W^{\mu-1}}{(T\,W+\vert\,x\,\vert)^{\mu}}\,\,,
\end{equation}
with $2<\mu<3$. In fact, the theoretical work of Ref.\cite{ANNA}
proves that the maximization of the Tsallis entropy\cite{COSTA} with
entropic index $q = 1 + 1/\mu$ yields for $\Pi$ an inverse power law
form with index $\mu$, provided that a transition of length
$|x|$ is supposed to be related to the time $t$ by
$|x| = Wt$.The adoption of this entropic argument changes (\ref{eq4}) into:
\begin{eqnarray}\label{eq7}
\frac{\partial}{\partial\,t}\sigma(x,t)&=&\frac{(\mu-1)\,(T\,W)^{\mu-1}
}{T}\Bigg[\int_{-\infty}^{\infty}
\frac{\sigma(x',t)\,dx'}{(T\,W+\vert\,x-x'\,\vert)^{2+\beta}}\nonumber\\&-&
\int_{-\infty}^{\infty}\frac{dx'}{(TW+\vert\,x'\,\vert)^{2+\beta}}\,\sigma(x,
t)\Bigg],
\end{eqnarray}
where $\mu=2+\beta$ with $0<\beta<1$. As shown in Appendix, the
Fourier transform analysis of (\ref{eq7}) proves  that in the
asymptotic regime ($k\,W\,T\ll 1$) this equation is equivalent to
the process of L\'evy diffusion of Eq.(\ref{final}) where   the
parameter $b$ reads:
\begin{equation} \label{b}
b=2\cos\big(\frac{\pi\alpha}{2}\big)
\Gamma(1-\alpha)\frac{(W\,T)^\alpha}{T}.
\end{equation}
In other words, we derive a result equivalent  to that of Section
II. Note that in the limiting case $T = 0$ the term within the
square brackets of Eq.(\ref{eq7}) becomes the
regularized\cite{gelfand} form of:
\begin{equation}
\int_{-\infty}^{+\infty} \frac{1}{|x-x^{\prime}|^{2 +
\beta}}\sigma(x^{\prime},t)dx^{\prime}
\label{toregularize}
\end{equation}
and, in this sense, coincides with the expression found by
Seshadri and West\cite{seshadriwest}. Note that keeping $T>0$
makes it possible for Eq.(\ref{eq7}) to cross the critical
condition $\beta = 1$ without meeting the divergence corresponding
to the L\'{e}vy prescription, namely the divergence of $b$ of
Eq.(\ref{defineb}) at $\alpha = 2$. The result of Eq.(\ref{eq7})
can be used also to study the region $\beta > 1$ corresponding to
the attraction basin of the ordinary central limit theorem.

It is important to observe that
the Markovian master equation here under study can
 be derived from the generalized
master equation of
Eq.(\ref{generalized}) by using the Markov condition:

\begin{equation}
\Pi(x) = \int_{0}^{\infty} \pi(x,t')dt'.
\label{markovapproximation}
\end{equation}
This is an axpect of crucial importance. In fact
Eq.(\ref{markovapproximation}) yields Eq. (\ref{eq6}), showing that the
Markov property makes it
possible to establish an important connection between dynamics and
thermodynamics.
The Markov perspective adopted in this Section is essential to
establish the key connection between the structure of Eq.(\ref{eq6}),
resulting from the adoption of entropic arguments, and the structure
of Eq.(\ref{eq1}), generated by the adoption of the renormalization
group arguments, which, in turn,  reflect genuinely Hamiltonian
properties \cite{ZAS}.
Note that we refer to Eq.(\ref{markovapproximation}) as a \emph{Markov
condition} rather than as a \emph{Markov approximation}. This is so
because the term ``approximation'' suggests a given departure from
the exact solution or, in other words, an error whose intensity must
be defined. We see, on the contrary, that the asymptotic regime of
Eq.(\ref{eq7}) coincides with that of Eq.(\ref{generalized}), if, the
latter equation is supplemented by the crucial condition of
Eq.(\ref{flight}), mirroring the dynamics illustrated in the
Introduction.

\section{The exact diffusion equation and the Markov regime}
Is there a connection between Eq.(\ref{eq3}) and Eq.(\ref{eq7})?
It is evident that this connection would be established by the
exact solution of (\ref{eq3}), if this yielded, in the asymptotic
time limit,  a diffusion process of L\'evy kind. It has to pointed
out, however,  that finding  an exact solution of (\ref{eq3}) is
not easy. In literature we find only solutions of (\ref{eq3})
based on approximations\cite{ALLEGRO,TC1,TC2}. In
Ref.\cite{ALLEGRO} a solution of Eq.(\ref{eq3}) was found, with a
Markov character, and corresponding to Eq.(\ref{eq7}) with only
the first of the two terms between the square brackets on the
\emph{r.h.s} of this equation. In ref.\cite{TC1} a non-Markov
solution has been discussed, which has been later judged to be the
correct solution\cite{TC2}.

The purpose of this Section is that of discussing these two proposed solutions
in the light of the  sporadic
randomness illustrated in Section I.
In the work of \cite{TC1} it was argued that  an analytical solution of
Eq. (\ref{eq3}) can be found, based on the fact that
a fractional derivative in time emerges from the r.h.s of (\ref{eq3})
if  the correlation function
$\Phi_{\xi}(t)$, which should fulfill the normalization
condition $\Phi_{{\xi}}(0) = 1$, is replaced by a function like
$const/t^{\beta}$. This means a function with the same long-time
property as the original correlation function, breaking
however the normalization condition at
$t=0$. This approximation results
in very appealing mathematical properties.
In fact, it has the nice effect
of resulting in a process with infinite memory, and in an analytical
expression for the effects that this infinitely extended memory has
on diffusion.
The adoption of this approximation yields
\cite{TC2} the rescaling of Eq.(\ref{eq9}).
It is straightforward to prove that this  rescaling can be obtained
from the Fourier-Laplace transform
of Eq. (\ref{eq3})
\begin{equation}
\hat{\sigma}(k,s) = \frac{<\xi^{2}>}{s + \hat{\Phi_{\xi}}(s) k^{{2}}},
\label{laplacefourier}
\end{equation}
where $\hat{\Phi}_{\xi}(s)$ is the Laplace transform of the
correlation function $\Phi_{\xi}(t)$. To derive the rescaling of
Eq. (\ref{eq9}) we have to replace in Eq. (\ref{laplacefourier})
$\hat{\Phi_{\xi}}(s)$ with  $s^{\beta-1}$ (see \cite{RG}). This
rescaling, however, conflicts with the numerical observation of
Ref.\cite{ALLEGRO}, which results in a different rescaling,
corresponding to that of Eq.(\ref{eq10}). The discussion of
Section II sheds light on the origin of this rescaling, different
from that of Eq. (\ref{eq9}). It seems to be evident to us that
the study of the asymptotic properties of Eq.
(\ref{laplacefourier}) resting on the limiting condition
$\lim_{s\rightarrow 0} \hat{\Phi}_{\xi}(s) = const \,s^{\beta-1}$,
loses any dependence on the key parameter $T$, and with it, on the
fact that there exists a propagation front moving with finite
velocity. This explains why the same method of time asymptotic
analysis applied to Eq. (\ref{nonmarkovapproximation}) yields the
correct rescaling. This is so because in this case the L\'{e}vy
walk nature of the process under study is retained by the kernel
$\kappa(x,t)$ due to the wise choice made for $\pi(x,t)$ in Eq.
(\ref{flight}).

In conclusion, we are convinced that the solution implying the
existence of an infinitely extended memory conflicts with
the numerical treatment of the diffusion process
resulting from the fluctuations of the dichotomous variable
$\xi$ with a non integrable correlation function
$\Phi_{\xi}(t)$. This is so because the steady action of sporadic
randomness has the effect of producing
a Markov statistics, although this occurs in the long-time limit.
How to make the rescaling of the
central part of the distribution $\sigma(t)$  to become  compatible
with the effect of sporadic randomness and with the predictions of the
L\'{e}vy-Gnedenko theorem\cite{gnedenko} in the long-time limit?
The most direct way to realize the correct rescaling
of the central part of the distribution and to make the Markov
property emerge is that of assuming that:
\begin{equation}\label{eq11}
\sigma(x,t-t')=\frac{1}{2}\int_{-\infty}^{\infty}
\delta(W\,t'-\,\vert\,x-x'\,\vert)
\,\sigma(x',t)\,dx'\,\,.
\end{equation}
As earlier pointed out, if we apply this condition to the term on
the \emph{r.h.s.} of Eq.(\ref{eq3}), we obtain an equation of
motion identical to one that would result from Eq.(\ref{eq7})
cancelling the second of the two terms within the square brackets
of this equation. This interesting result implies some algebra
based on the method of integration by parts and the properties of
the delta of Dirac. More details on this calculation can be found
in Ref.\cite{ALLEGRO}.

It is important to point out the physical meaning of the
constraint of Eq.(\ref{eq11}). This means that we imagine a
condition still unaffected by randomness, since this constraint
would be rigorously valid only in the case of merely ballistic
motion. Yet, the effect of replacing  (\ref{eq11}) into the r.h.s
of (\ref{eq3}) is that of producing a Markov structure as an
effect of  carrying out the integration on $t'$ in the time
convoluted form of Eq.(\ref{eq3}). We are convinced that the
emergence of this Markovian structure is not an artefact of the
approximation of Eq.(\ref{eq11}). The error associated to this
approximation is not the emergence of the Markov structure. This
error is totally different in nature, and can be easily evaluated.
In fact, as  repeatedly pointed out earlier, this approximation
has the effect of resulting only in the first term on the r.h.s of
(\ref{eq7}). The error associated   to this approximation is
signalled  by the breaking of the norm conservation. It is evident
in fact that the condition:
\begin{equation}\label{12}
\int_{-\infty}^{\infty}\sigma(x,t)\,dx=1
\end{equation}
is fulfilled by both (\ref{eq3})  and (\ref{eq7})
and that in the latter case this is the
consequence of the wise structure of the master
equation of (\ref{eq4}). This is the reason why
we look at the master equation of (\ref{eq7})
as a natural bridge to cross when moving from
the dynamical perspective of (\ref{eq3}) to the
final regime of L\'evy kind. As noticed
in Section III, the emergence of
this final condition from (\ref{eq7}) is made
 by
using the mathematical approach of the Appendix which proves that
the Fourier transform of (\ref{eq7}) yields the genuinely L\'{e}vy
process of diffusion of Eq.(\ref{final}). We note that in the
earlier work\cite{ALLEGRO}, \cite{hope} no proper attention was
devoted to the crucial fact that the second term within the square
bracket of Eq.(\ref{eq7}) is essential for a proper derivation of
the L\'{e}vy processes, or of the equivalent fractional
derivative.

\section{Concluding remarks}
In conclusion, we have provided a convincing demonstration of how to
derive L\'{e}vy processes from within a dynamic approach.
As pointed out in an earlier work \cite{hope} the L\'{e}vy process
corresponds to a form of fractional calculus which can be regarded
as a form of macroscopic manifestation of microscopic randomness.
However, it seems to us that fractional derivatives in time have a
different meaning from fractional derivatives in space. The choice
made by the authors of Ref.\cite{TC2} has the effect of relating
the solution of Eq.(\ref{eq3}) to a form of fractional derivative in
time. It seems to us that this choice corresponds to a case where
the decay of the correlation function is not originated by the
sporadic action of randomness on a single trajectory. In this latter
case, as shown in this paper,
to obtain the correct result we are forced to make the
Markov property emerge, and this is
realized by adopting the trick of Eq.(\ref{eq11}).
The assumption made by the authors of Ref.\cite{TC2}, on the contrary,
seems
to imply that the
decay of the correlation function is originated by
 a  statistical distribution
over a range of initial conditions. Whether or not this is
compatible with the dichotomous nature of the diffusion generator
is not quite clear to us.

Should the proof be given that the relaxation of the correlation
function $\Phi_{\xi}$ can also be determined by statistics as well as
by dynamics,
with no conflict with the dichotomous
property behind Eq. (\ref{eq3}), we would reach the impressive
conclusion that this exact equation admits two distinct classes of
solution, determined by the extra information about
the physical origin of the relaxation process. We are inclined to believe
that the relaxation of a dichotomous
variable can only be compatible with
the action of a sporadic randomness. Thus the
emergency of the Markov property at extremely long time is a
fair reflection of the dynamical processes generating the L\'{e}vy
diffusion processes, not to speak of the fact that the corresponding
rescaling of the central part of the distribution
fits the results of the numerical calculations\cite{ALLEGRO}.

It is also important to stress that the repeated action of randomness
is subtly related to the possibility of establishing a connection
between mechanics and thermodynamics also in the case, studied in
this paper, of apparently
infinite memory. The random seed is given by the fact that the
duration of the times of sojourn in the laminar region cannot be
predicted. This is so because of the random injection into the
laminar region from the chaotic part of the map \cite{IM}.
This is the reason why we have to introduce probabilistic arguments
within the dynamical picture of the process under study. This is also
the reason why, as shown by\cite{ANNA}, the shape of the density
distribution of
Eq.(\ref{eq1}) can be predicted by using entropic arguments, provided
that the non-extensive form of entropy advocated by Tsallis
\cite{COSTA} is used. In other words, both the adoption of entropic
arguments and the birth of L\'{e}vy statistics rest on the emergence
of probabilistic aspects generated by a sporadic form of randomness.

Finally, we are left with the intriguing issue of the dependence
of all these properties on the space dimensions. The treatment of
this paper has been confined to the one-dimensional case, where
the crucial action of the stability islands pointed out by the
theoretical analysis of Zaslavski\cite{ZAS} is correctly mirrored
by the dichotomous nature of the fluctuating variable $\xi$. We
are convinced that the essence of the present treatment can be
extended to the multidimensiuonal case, including the more
realistic three dimensional case. However, we have to recognize
that this extension is not a trivial matter and that further
research work has to be done to settle the technical problems
triggered by the two- and three-dimensional case. ${}\\$

\appendix{\textbf{APPENDIX}}
${}\\$

We define the function

\begin{equation} \label{exp}
D^{\alpha}_{x} e^{x}=\sum
^{\infty}_{n=0}\frac{x^{n-\alpha}}{\Gamma(n+1-\alpha)}\equiv
E^{x}_{\alpha}
\end{equation}
This function is a generalization of the exponential function. In
turn, this generalized exponential is used to derive the following
form of  generalized trigonometric function:

\begin{equation} \label{trigo1}
sin_{\alpha}x=\frac{E^{\imath x}_{\alpha}-E^{-\imath
x}_{\alpha}}{2\imath }
\\
\end{equation}
and
\begin{equation} \label{trigo2}
cos_{\alpha}x=\frac{E^{\imath x}_{\alpha}+E^{-\imath
x}_{\alpha}}{2}
\end{equation}
The corresponding expansion in a power series are

\begin{equation} \label{serie1}
   sin_{\alpha}x=
\sum^{\infty}_{n=0}\frac{x^{n-\alpha}\sin[(n-\alpha)\pi/2]}{\Gamma(n+1-\alpha)}
\\
\end{equation}
and
\begin{equation} \label{serie2}
     cos_{\alpha}x=\sum^{\infty}_{n=0}\frac{x^{n-\alpha}\cos[(n-\alpha)\pi/2
]}{\Gamma(n+1-\alpha)} . \\
\\
\end{equation}
Note that the functions defined in Eq.(\ref{exp}) fulfill the
important relation
\begin{equation}\label{integr}
\int x^{\alpha} e^{x}dx=\Gamma(1+\alpha)e^{-x}E^{x}_{-1-\alpha}.
\end{equation}

We are now ready to address the problem raised by  Eq.(\ref{eq7}),
namely,the evaluation of the Fourier transform of the function
$1/(TW+\vert\,x-x'\,\vert)^{2+\beta}$. For this purpose we can use
Eq.(\ref{integr}) and we find:
\begin{eqnarray}\label{fourier}
\int\limits_{-\infty }^{+\infty }\frac{e^{\imath
kx}}{(a+\vert\,x\,\vert)^{2+\beta}}dx&=&
2Re\int\limits_{0}^{+\infty}
\frac{e^{ikx}dx}{(a+|x|)^{1+\alpha}}\nonumber\\
 &=&f(k).
\end{eqnarray}
On the other hand:
\begin{eqnarray} \label{fk}
f(k)&=& 2 \int\limits_{0 }^{+\infty }\frac{\cos
kx}{(a+x)^{1+\alpha}}dx\nonumber\\ &=& 2\Gamma \left(-\alpha
\right) |k|^{\alpha } \Big[\,sin \big[\frac{\pi}{2}(1+\alpha)
+\vert\, ka\,\vert\,\big]\nonumber \\ &&\qquad
-\tilde{sin}_{\alpha}\big[\frac{\pi}{2}(1+\alpha) +\vert\,
ka\,\vert \,\big]\,\Big],
\end{eqnarray}
where $a\equiv WT$ and

\begin{eqnarray}\label{solut}
\tilde{\sin}_{\alpha }\big[\frac{\pi}{2}(1+\alpha) +\vert\,
ka\,\vert \,\big] = cos\big[\frac{\pi}{2}(\alpha +1)\,\big]
sin_{\alpha}\vert\,ka\,\vert\nonumber \\ +
sin\big[\frac{\pi}{2}(\alpha +1)\big] cos_{\alpha}\vert\,ka\,\vert
.
\end{eqnarray}

We are interested in the case $ka\rightarrow 0$. Thus we use
(\ref{serie1}), (\ref{serie2}) and (\ref{solut}) to evaluate the
Fourier transform of interest keeping only the leading vanishing
and diverging terms. We thus obtain:
\begin{eqnarray}\label{approx}
\int_{-\infty}^{+\infty}dx \frac{exp(ikx)}{(a+|x|)^{(1+\alpha)}}
 \approx 2 |k|^{\alpha}\Gamma(-\alpha)\Big[\cos(\frac{\pi
 \alpha}{2})\nonumber\\
                                      - \sin(\frac{\pi
                                      \alpha}{2})\vert\,ka\,\vert
                                      -\frac{\vert\,ka\,\vert^{-\alpha}}{\Gamma(1-\alpha)}
+\frac{\vert\,ka\,\vert^{2-\alpha}}{\Gamma(3-\alpha)} \Big]
                                       \end{eqnarray}
 In conclusion,the Fourier transform of (\ref{eq7}) yields
 \begin{equation}
\frac{\partial} {\partial\,t}\,\hat{\sigma}(k,t)=  c[f(k) -f(0)]
\hat{\sigma} (k,t), \label{almostfinal}
\end{equation}
where $f(k)$ is given by Eq.(\ref{fk}) and
\begin{equation}
c \equiv \frac{1}{T}(\mu-1)T^{(\mu-1)}
W^{(\mu-1)}=\frac{1}{T}(\mu-1)a^{(\mu-1)}. \label{cdefinition}
\end{equation}
where $\mu -1 = \alpha$. After some algebra we find that $f(0) =
2/(\alpha\,a^\alpha)$ end the final coefficient in front of
$\sigma(k,t)$ is:
\begin{eqnarray}
2\cos(\frac{\pi \alpha}{2})\Gamma(-\alpha)\frac{\alpha\mid
ka\mid^{\alpha}} {T}&=&-2\cos(\frac{\pi
\alpha}{2})\Gamma(1-\alpha) \frac{a^{\alpha}}{T}\mid
k\mid^{\alpha}. \nonumber \\
\label{coefficient}
\end{eqnarray} As a final result we get Eq.(\ref{final}).

\end{document}